\documentclass[12pt]{book}

\usepackage[dvips]{graphicx,color}
\usepackage{makeidx,tsukuba}

\makeauthorindex
\makeindex

\begin{document}

\BookTitle{\itshape The 28th International Cosmic Ray Conference}
\CopyRight{\copyright 2003 by Universal Academy Press, Inc.}
\pagenumbering{arabic}

\chapter{
Identification of Showers with Cores Outside the ARGO-YBJ Detector}

\author{
%
%
G. Di Sciascio,$^1$ C. Bleve,$^2$ T. Di Girolamo,$^1$ D. Martello,$^2$, E. Rossi,$^1$ 
for the ARGO-YBJ Collaboration [3]\\
{\it (1) Dip. di Fisica Universit\'a di Napoli and INFN sez. di Napoli, Napoli, 
Italy\\
(2) Dip. di Fisica Universit\'a di Lecce and INFN sez. di Lecce, Lecce, Italy} \\
}

\section*{Abstract}
In any EAS array, the rejection of events with shower cores outside the detector boundaries is of 
great importance. A large difference between the true and the reconstructed shower core positions 
may lead to a systematic miscalculation of some shower characteristics. Moreover, an accurate 
determination of the shower core position for selected internal events is important to reconstruct
the primary direction using conical fits to the shower front, improving the detector angular 
resolution, or to performe an efficient gamma/hadron discrimination.

In this paper we present a procedure able to identify and reject showers with cores outside the 
ARGO-YBJ carpet boundaries. 
A comparison of the results for gamma and proton induced showers is reported.

\section{Introduction}
Showers of sufficiently large size can trigger a detector even if their core is located 
outside its boundaries. 
The corresponding core positions are generally reconstructed not only
near the carpet edges but also well inside the boundaries.
As a consequence, sofisticated algorithms able to reduce the contamination of external events 
are needed.
The goal is to identify and reject a large fraction of external events before 
exploiting any reconstruction algorithm, only by using some suitable parameters.

In this paper we present a reconstruction procedure able to identify and reject a large fraction
of showers with cores outside the ARGO-YBJ detector.

\section{Identification of external events}

The ARGO-YBJ detector consists of a single layer of RPCs with dimensions of
$\sim 74\times 78~m^2$. The area surrounding this central detector ({\it carpet}), up to 
$\sim$ 100 $\times$ 110 $m^2$, is partially ($\sim 50 \%$) instrumented with RPCs 
({\it guard-ring}).
The basic element is the logical {\it pad} ($56\times 62~cm^2$) which defines the time and space 
granularity of the detector. The detector is divided in 6 $\times$ 2-RPC units (clusters): 
the central carpet contains 10 $\times$ 13 clusters.
For a detailed description of the ARGO-YBJ detector see [3].

Various parameters based on particle density or time information are under investigation to 
identify
showers with core position outside a given fiducial area. The most interesting ones are 
the following: (1) position of the cluster with the highest particle density, 
(2) position of the cluster row/column with the highest total particle density,
(3) mean distance $R_p$ of all fired pads to the reconstructed shower core.

To perform these calculations we have simulated, via the Corsika code [1], gamma and proton 
induced showers 
with energy spectra ($\sim E^{-2.5}$ and  $\sim E^{-2.75}$, respectively) ranging from 100 GeV to 
50 TeV. The detector response has been simulated via a GEANT-3 based code.

As an example, in Fig.\,1 we show the distributions of the position of the cluster 
with the highest particle density for $\gamma$-induced showers.
In the plots we compare the events with the core really external to a 80 $\times$ 80 
m$^2$ fiducial area (solid histograms) and the truly internal ones (dashed histograms).
To investigate the discrimination power of this particular parameter we have simulated
a detector completely 
instrumented up to $\sim$ 100 $\times$ 110 $m^2$, i.e., containing 14 $\times$ 17 clusters.
Therefore, the cluster coordinates run from 1 to 14 (X view) and from 1 to 17 (Y view) 
starting from the lower left corner of the carpet. 

\begin{figure}[t]
\vfill \begin{minipage}[t]{.47\linewidth}
  \begin{center}
    \includegraphics[height=17pc]{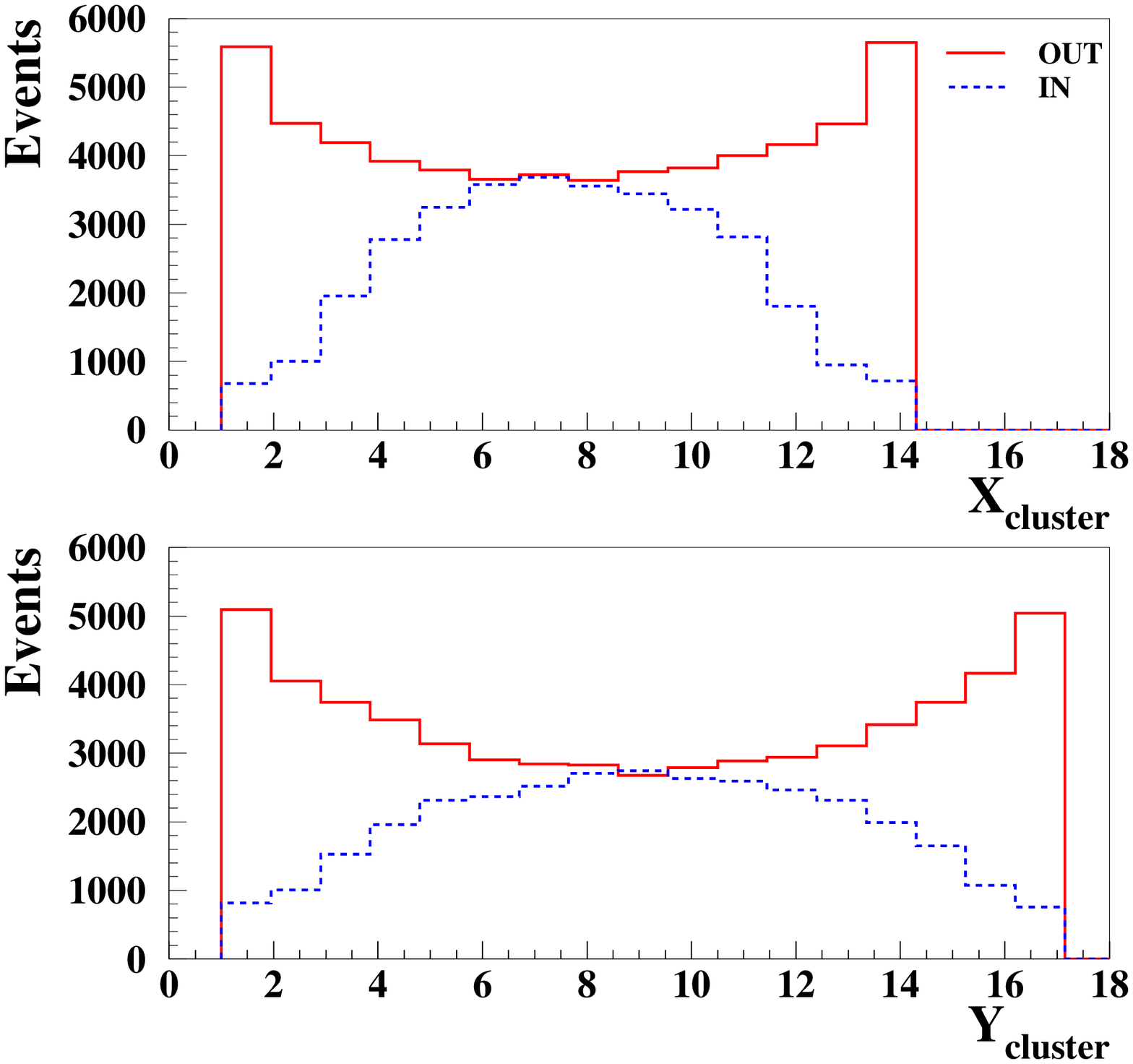}
  \end{center}
  \vspace{-0.5pc}
    \caption{Coordinate distributions of the cluster with the highest particle density 
for $\gamma$-induced events with a pad multiplicity $N_{hit} > 100$.}
\end{minipage}\hfill
\hspace{-0.5cm}
\begin{minipage}[t]{.47\linewidth}
  \begin{center}
    \includegraphics[height=17pc]{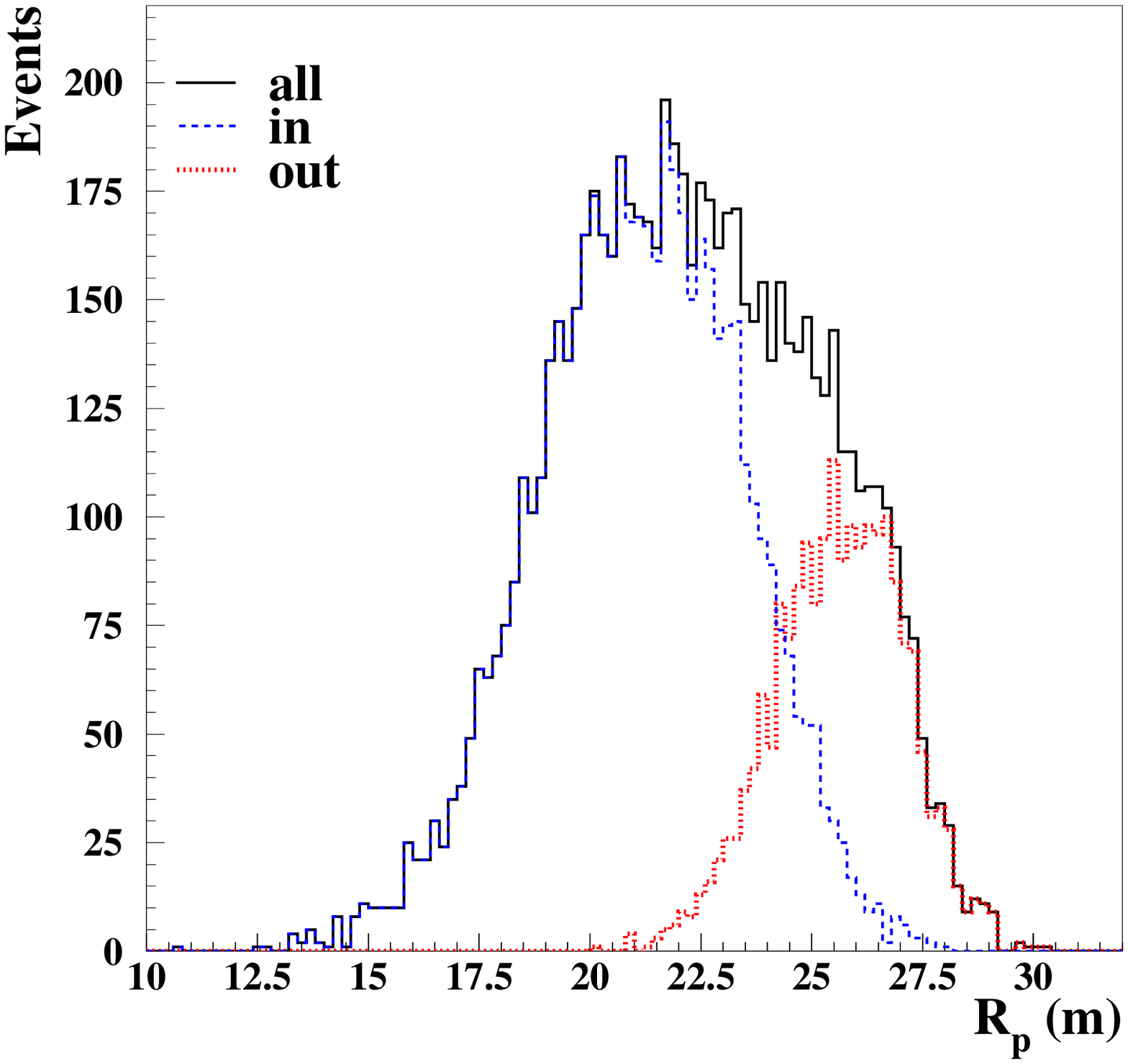}
  \end{center}
  \vspace{-0.5pc}
    \caption{ Distributions of the parameter $R_p$ (solid histograms) for IN 
reconstructed showers, for $\gamma$-induced events ($N_{hit} > 100$).}
\end{minipage}\hfill
\end{figure}

The $R_p$ distribution for showers reconstructed inside a 80 $\times$ 80 m$^2$
fiducial area is shown in Fig.\,2 (solid histogram). 
The dashed line refers to truly IN events while the dotted histogram refers to OUT
showers erroneously reconstructed as internal. 
The shower cores have been calculated by means of the simple center of gravity method.
As can be seen, the parameter $R_p$ identifies quite well the events with core outside 
the carpet. Large distances between the truly and the reconstructed shower axis lead to larger 
$R_p$ values. This fact offers the possibility to define a cut in $R_p$ to 
identify these events. A conservative choice is to reject showers with $R_p > 25$ m. 

From these studies it follows that the identification of a large fraction of external events 
can be achieved by defining a suitable fiducial area togheter with a combination of 
cuts in the parameters discussed above.

\section{Maximum Likelihood Method ({\tt LLF}) }

Different algorithms have been investigated to reconstruct the shower core position in the
ARGO-YBJ experiment [2]. The most performant is the Maximum Likelihood Method.
We point out that expression for -LLF of [2] refers to the case of a Poisson distribution in 
which the pads are not fired with 
probability $P_i (0)$ or fired with probability $P_i (>0) = 1 - P_i (0)$ (hereafter
'{\tt LLF1} method'). 
In our study almost always the fired pads have particle multiplicity 1, and therefore such a 
simple discrimination can be made. However, if we consider a larger area as the
whole RPC, the multiplicity can be $>$ 1, and the proper Poisson
distribution on the fired RPCs appears more adequate. In this case the sum on fired elements is:
\begin{equation}
-\Sigma_j ln P_j (>0) = -\Sigma_j N_j ln(\rho_j) - ln(S_{RPC}) \Sigma_j N_j
 +\Sigma_j ln(N_j !) + S_{RPC} \Sigma_j \rho_j
\end{equation}
where $N_e\cdot \rho_j$ is the particle density expected on the j-th RPC at a distance $R_j$ 
from the core, $N_j$ is the recorded particle number and $S_{RPC}$ is the RPC area. 
The shower size can be calculated via the equation
\begin{equation}
 N_e =  \frac{\Sigma_j N_j } {S_{RPC}\Sigma_j \rho_j}.
\end{equation}
We define this calculation the '{\tt LLF2} method'.
As a consequence, we expect that the differences between {\tt LLF1} and {\tt LLF2}
increase with the particle density, for a fixed area.
In Fig.\,3 we compare the shower core position resolution calculated by applying 
the {\tt LLF1} and {\tt LLF2} methods on the RPCs for $\gamma$-induced showers with the core
randomly sampled inside a 80 $\times$ 80 m$^2$ area. 
As expected, the resolution worsens with multiplicity if the {\tt LLF1} approach is applied 
when the number of particles hitting the RPC is $>$ 1.
We note that for very low multiplicities ($N_{hit} < 80$) the method {\tt LLF1} is more 
performant than {\tt LLF2}. In fact, the algorithm based on RPC occupancy ({\tt LLF1})
provides a better representation of the hit distribution in very poor showers.

For very high multiplicities ($N_{hit} > 10^3$) the shower core position 
is determined by {\tt LLF2} with an uncertainty $<$ 1 m.

\section{Results }

A possible procedure to reject external events in the ARGO-YBJ detector is the following one:
(1) Rejection of the events whose highest density clusters are on the guard ring 
(X = 1, 14; Y = 1, 17) or on the boundaries of the central carpet (X = 3, 12; Y = 3, 15).
(2) Rejection of the events whose highest total density rows or columns are respectively 
in positions $\{1,3,15,17\}$ or $\{1,3,12,14\}$.
(3) Reconstruction of core coordinates $\{ X_c,Y_c \}$ using the Maximum Likelihood Method.
(4) Further rejection of events with $R_p > 25$ m.
\begin{figure}[t]
\vfill \begin{minipage}[t]{.47\linewidth}
  \begin{center}
    \includegraphics[height=17pc]{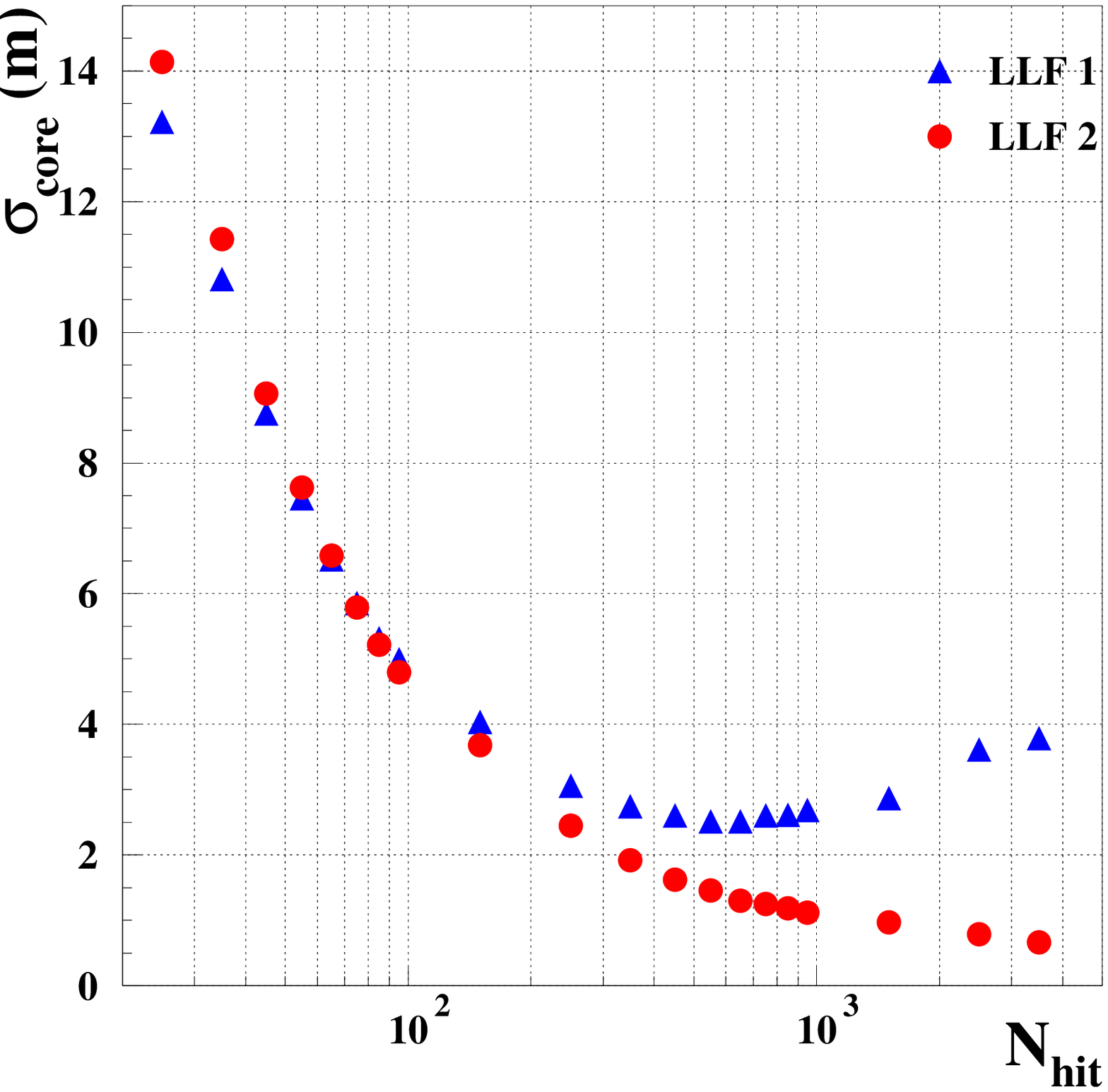}
  \end{center}
  \vspace{-0.5pc}
    \caption{Comparison between the shower core position resolutions obtained using 
LLF1 and LLF2 methods.} 
\end{minipage}\hfill
\hspace{-0.5cm}
\begin{minipage}[t]{.47\linewidth}
  \begin{center}
    \includegraphics[height=17pc]{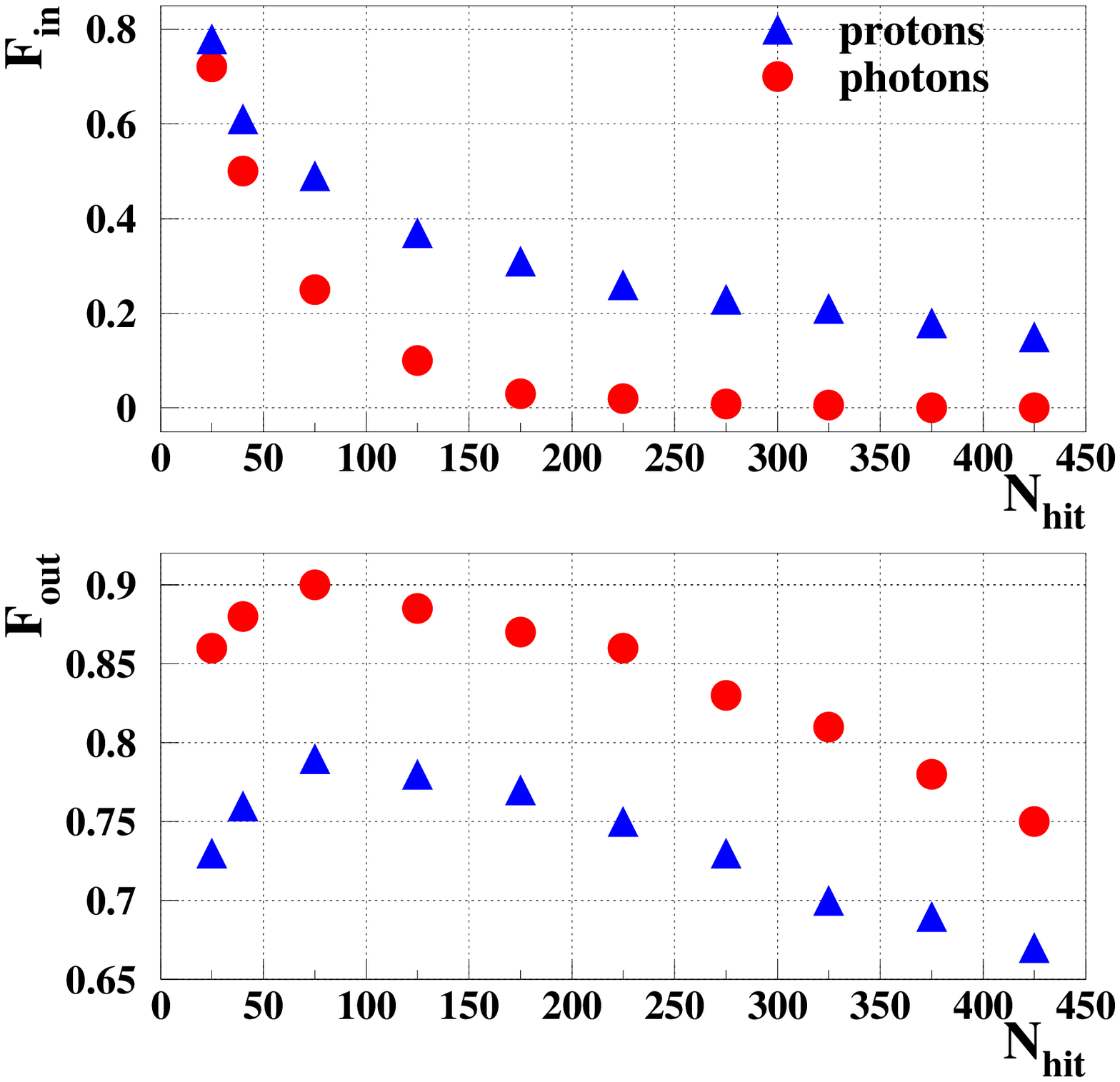}
  \end{center}
  \vspace{-0.5pc}
    \caption{Fraction of truly internal and external events rejected by the selection 
procedure (1) - (4).}
\end{minipage}\hfill
\end{figure}

In Fig.\,4 the fraction of events (internal and external to an area of 
80 $\times$ 80 m$^2$, respectively) rejected after the steps (1) - (4) is reported. 
As can be seen, this procedure is able to identify and reject a large fraction of external events.
For low multiplicities ($N_{hit} < 100$) a significative fraction of internal events is 
erroneously rejected, especially in proton-induced showers.

\section{References}

\vspace{\baselineskip}
\re
1.\ Heck D.\ et al.\ 1998, Report {\bf FZKA 6019} Forschungszentrum Karls\-ruhe.
\re
2.\ Martello D., Bleve C., Di Sciascio G.\ 2001, 27th ICRC, Hamburg, 7, 2927.
\re
3.\ Surdo A.\ et al.\ 2003, in this proceedings.

\endofpaper
\end{document}